\pdfoutput=1
\documentclass{JINST}

\usepackage{amsmath}

\title{Measurement of the drift field in the ARGONTUBE LAr TPC with 266~nm pulsed laser beams}

\author{ A.~Ereditato, D.~Goeldi, S.~Janos, I.~Kreslo\thanks{Corresponding author.}~, 
M.~Luethi, C.~Rudolf~von~Rohr, M.~Schenk, T.~Strauss, M.~S.~Weber and M.~Zeller\\
\llap Albert Einstein Center for Fundamental Physics, Laboratory for High Energy Physics\\
  Universit\"{a}t Bern, Switzerland\\
E-mail: \email{igor.kreslo@lhep.unibe.ch}}

\abstract{ARGONTUBE is a liquid argon time projection chamber (LAr TPC) with a drift field generated in-situ by a Greinacher voltage multiplier circuit. We present results on the measurement of the drift-field distribution inside ARGONTUBE using straight ionization tracks generated by an intense UV laser beam. Our analysis is based on a simplified model of the charging
of a multi-stage Greinacher circuit to describe the voltages on the field cage rings.}


\begin{document}
\section{Introduction}
The calibration of the electric drift field in large volume liquid argon time projection chambers (LAr TPC) is required to account and correct for field disuniformities that can be mainly caused by two different mechanisms. First, the sensitive volume of a detector located underground at a shallow depth or at ground surface is subject to permanent ionization by cosmic rays, leading to the accumulation of positive argon ions in the detector drift region. These ions can produce a substantial electrostatic field that distorts the initially uniform drift field, hence affecting the reconstruction of particle tracks. The deviation of the reconstructed ionization track coordinates from the true ones can be as large as $\approx$10~cm after a 2.5~m drift distance\footnote{LAr TPC for MicroBooNE experiment, www-microboone.fnal.gov}. A second cause of field disuniformities originates from an initial non-uniform potential distribution over the field-shaping rings in the drift region. 

Depending on the aspect ratio (drift length to field cage width) of the detector, one of the two distortion causes can dominate over the other. Building an accurate three-dimensional map of the resulting electric field in the detector allows to efficiently compensate for these distortions and to reconstruct the true track geometry.

The map of the drift field can be derived by means of straight ionization tracks covering the entire sensitive detector volume, $e.g.$\ induced by high energy cosmic ray muons. However, to achieve the required granularity of the field map, a sufficiently large statistics is needed. Furthermore, the larger is the detector, the smaller is the high energy fraction of the muon spectrum that can be used for this kind of analysis. The limitation originates from the significant Coulomb multiple scattering of muons in the relatively dense liquid argon medium. A method to produce a large number of straight ionization tracks in a controlled way which are neither subject to Coulomb multiple scattering nor to $\delta$-ray emission would significantly simplify the task of obtaining such a drift field map in large volume LAr TPCs.

The ARGONTUBE detector is a liquid argon time projection chamber that allowed to achieve for the first time a 5~m long drift distance for ionization electrons in liquid argon~\cite{ARGONTUBE0,ARGONTUBE1,ARGONTUBE2}. Given the high aspect ratio of about 25 for the detector sensitive volume (narrow field cage), the mirror charge induced on the surface of the field-shaping rings leads to a relatively fast removal of the positive ions. This  brings the distortion due to ion space charge to a negligibly low level.

Therefore, the dominating field distortion in ARGONTUBE originates from the non-uniform potential distribution on the field-shaping rings created by the undercharged high voltage generator. In order to demonstrate the feasibility of calibrating the drift field by means of straight ionization tracks, the detector is equipped with a high-power pulsed UV laser beam. The technique is based on the process of multi-photon ionization of argon atoms in liquid by a narrow ($\approx$ 2 mm diameter) beam of UV radiation with a wavelength of 266~nm~\cite{Badhrees:2010zz}.

\section{Drift field in ARGONTUBE}
\begin{figure}[ht]
\centering	
\includegraphics[width=1\linewidth]{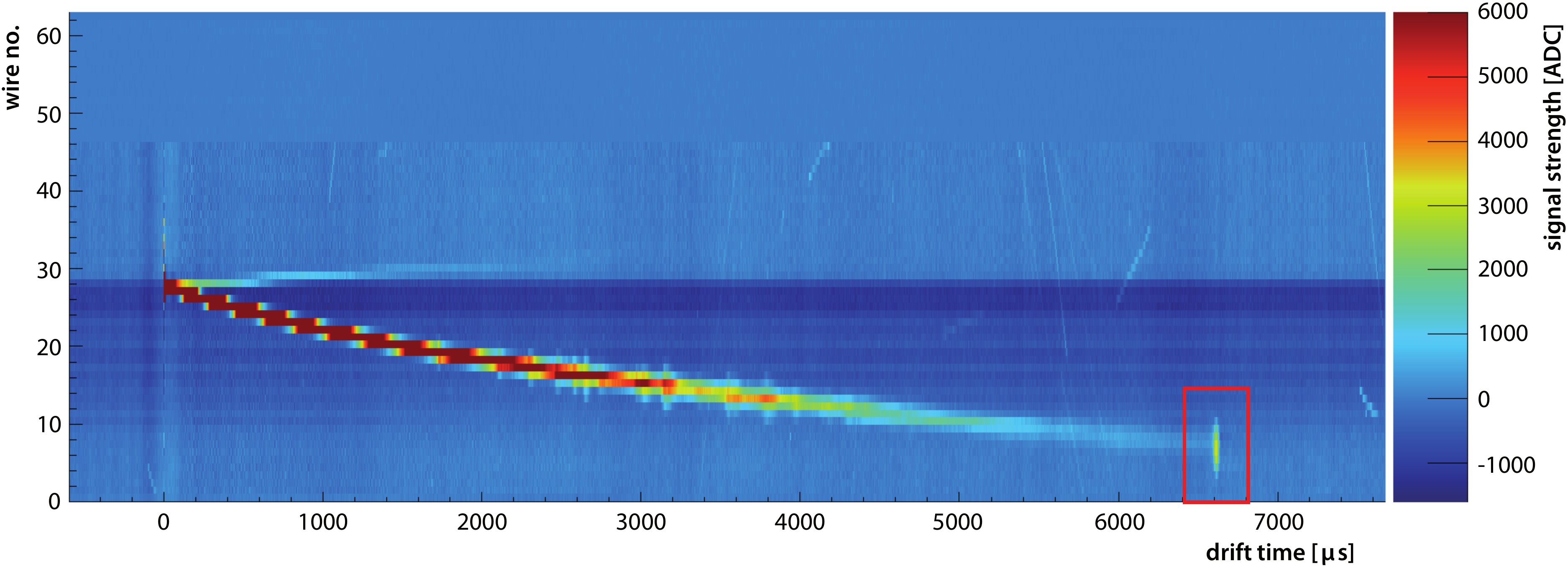}
\caption{A typical laser-induced ionization track in ARGONTUBE entering the detector from the top (left in the figure) and crossing the entire sensitive volume longitudinally. The red rectangle marks the electron cloud released from the detector cathode by photoelectric effect at about 5 mm drift distance.}
\label{fig:lastrack}
\end{figure}

For a detailed description of the ARGONTUBE detector and of its subsystems and, in particular, laser we refer to previous publications~\cite{ARGONTUBE0,ARGONTUBE1,ARGONTUBE2}. Electrostatic simulations of the TPC field cage indicated a high uniformity of the electric drift field within the sensitive volume for the chosen field cage geometry~\cite{ARGONTUBE2}. However, during detector runs it became evident that the drift field was not as uniform as expected from simulations. The ionization tracks produced by cosmic ray muons had apparent strong curvature at the readout plane. 
The curvature was not only much stronger than expected from Coulomb multiple scattering, but also systematic. A similar effect was observed for straight laser-induced ionization tracks which are free of Coulomb multiple scattering (Figure~\ref{fig:lastrack}). Such a behavior can be attributed to substantial longitudinal and transverse drift field distortions. 

\begin{figure}[ht]
\centering	
\includegraphics[width=0.85\linewidth]{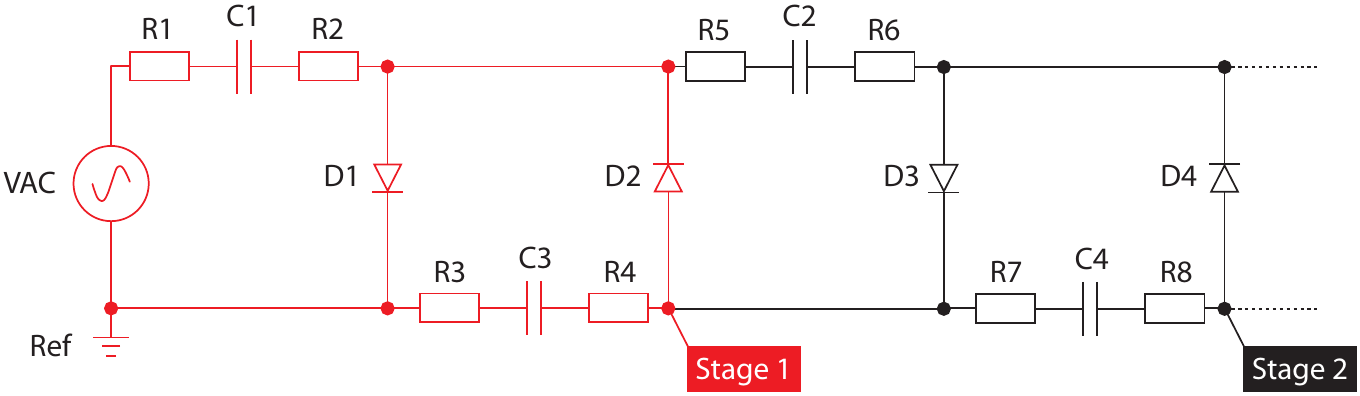}
\caption{Electric scheme showing the first two stages of the Greinacher circuit installed in ARGONTUBE. It consists of an alternate current voltage source (VAC), capacitors (C), diodes (D) and current-limiting resistors (R).}
\label{fig:scheme}
\end{figure}

\begin{figure}[ht]
\centering	
\includegraphics[width=0.6\linewidth]{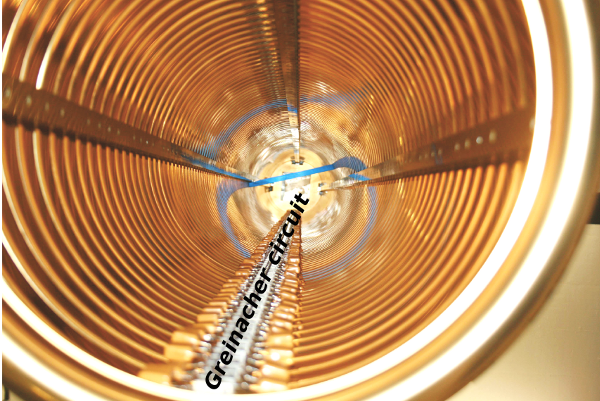}
\caption{The ARGONTUBE field cage seen from the top. The Greinacher circuit is mounted on the inner side of the field cage.}
\label{fig:grein}
\end{figure}

The high voltage needed to set up the electric drift field in ARGONTUBE is directly generated inside the cryostat by means of a Greinacher voltage multiplier \cite{Greinacher}. It is a circuit consisting of capacitors and diodes, which is driven by an alternate current voltage source (Figure~\ref{fig:scheme}). 119 multiplier stages are installed in ARGONTUBE to reach the required high voltage. The longitudinal component (w.r.t TPC drift direction) of the observed field distortions can be explained by the Greinacher multiplier being in an only partially charged state. Reaching fully charged state would take an infinite time. In the case of ARGONTUBE, the charging is stopped when the monitored input AC current drops down to noise level. The transverse parasitic component of the drift field, on the other hand, is mostly due to the fact that the Greinacher circuit is installed on the inner surface of the ARGONTUBE field cage (Figure~\ref{fig:grein}). 

\subsection{Model of the Greinacher voltage generator}
In the first approximation, the behavior of the potentials at the Greinacher circuit stages can be studied analytically. Let $N$ be the total number of stages of the multiplier and $U(n,\,t)$ the output voltage at stage $n\in[1,N]$ after a charging time $t$, and be $U(n,\,t\!=\!0) = 0\,\,\,\forall n$. Considering a single-stage circuit, $i.e.$\ $N=1$, the output voltage as a function of $t$ is given by 
\begin{equation}
  U(n\!=\!1,\,t) = U_\infty \left[1 - e^{-t/\tau_1}\right], \label{eq:SingleStage}
\end{equation} 
where $U_\infty$ denotes asymptotic value of the output voltage at the stage after an infinite charging time, and is referred to as the nominal voltage. $\tau_1$ is defined by  $R$ and $C$, the resistance and the capacitance of the single-stage circuit, .  
For a Greinacher multiplier with an arbitrary number $N$ of identical stages and a characteristic time constant $\tau$, the output voltage at stage $n$ is determined by replacing $\tau_1$ in Equation~\ref{eq:SingleStage} by $\tau_n$, to take into account all the resistive and capacitive components from stage $1$ to stage $n$. Since the individual stages are assumed to be identical, $\tau_n$ can be expressed as a fraction of $\tau$. Precise determination of  $\tau$ is not required for the presented analysis. The parameter, that defines how close is the circuit to its asymptotic state after charging time $t$ is defined by $t/\tau$ ratio.
\begin{equation}
 \tau_n =  \frac{n}{N} \; \tau.
\end{equation}
The potential distribution is then defined by
\begin{equation}
  U(n, \,t) = U_\infty^n \; \left[1 - \exp\left(-\frac{N}{n \tau}\;t\right)\right], \hspace{1cm} \text{where} \hspace{1cm} U_\infty^n = \frac{n}{N} \; U_\infty \label{eq:nStage}
\end{equation}
denotes the nominal voltage at stage $n$.

In ARGONTUBE one stage is added to the multiplier cascade for every additional field cage ring electrode. To describe the potential in a continuous way, we linearly interpolate the electric potential between each two consecutive rings along the drift spatial coordinate $z$ and replace  in Equation~\ref{eq:nStage} the integer numbers $n$ and $N$ by $z$ and $z_c$, respectively. 
The quantity $z_c$ corresponds to the location of the TPC cathode and $z=0$ defines the position of the charge readout plane. The function of the electric potential given in Equation~\ref{eq:nStage} is reformulated to
\begin{equation}
  U(z, \,t) = U_\infty \; \frac{z}{z_c} \; \left[1 - \exp\left(-\frac{z_c}{z \tau}\;t\right)\right], \label{eq:GreinModel}
\end{equation}
where $U_\infty$ can be interpreted as the cathode nominal voltage. Since the readout plane is connected to ground, $U(z\!=\!0, \,t) = 0\,\,\,\forall t$. 
The longitudinal electric field $E_L$ is given by
\begin{equation}
  E_L(z, \,t) = \frac{\partial U(z,\,t)}{\partial z} = \frac{U_\infty}{z_c} \left[1 - \exp\left(-\frac{z_c}{z \tau}\;t\right) - \frac{z_c t}{z \tau}\,\, \exp\left(-\frac{z_c}{z \tau}\;t\right)\right]. \label{eq:GreinModelE}
\end{equation}

\begin{figure}[ht]
  \centering
  \includegraphics[width=0.99\textwidth]{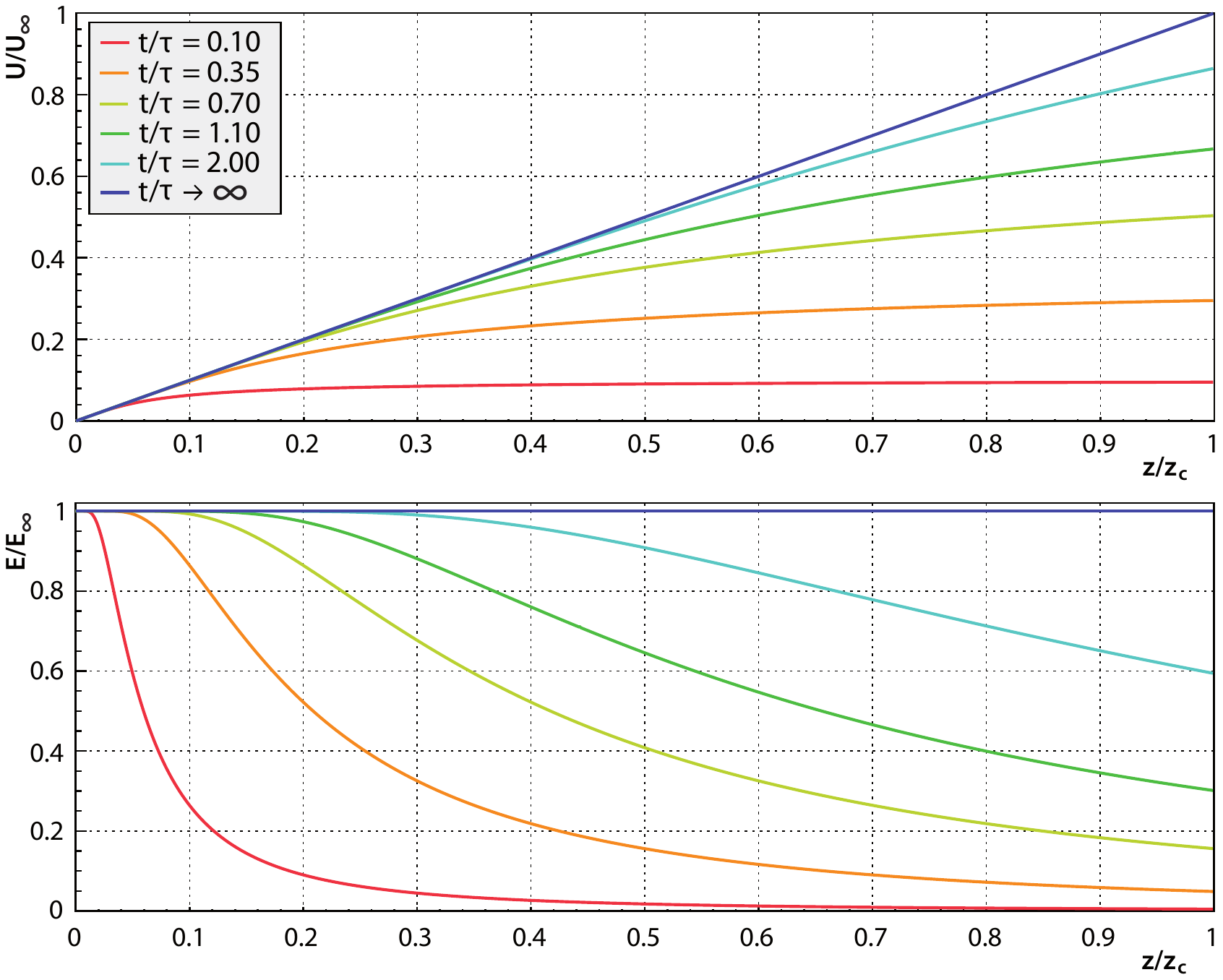}
  \caption[Illustrations of the 119 stages Greinacher model.]{Illustrations of the voltage (top) and the longitudinal electric field (bottom, here denoted by $E$) obtained with the Greinacher multiplier for different charging states $t/\tau$ as a function of the drift coordinate $z$. The graphs result from Equations~\ref{eq:GreinModel} and~\ref{eq:GreinModelE}. The dark blue curves correspond to a fully charged Greinacher circuit.}
   \label{fig:GreinacherModel}
\end{figure}

If $t\!\rightarrow\!\infty$, the exponential terms in Equations~\ref{eq:GreinModel} and~\ref{eq:GreinModelE} vanish and the potential follows a linear behavior $U(z, \,t\!\rightarrow\!\infty) = U_\infty\cdot z/z_c$ with respect to $z$, corresponding to a constant and uniform drift field $E_L(z, \,t\!\rightarrow\!\infty) = U_\infty/z_c$ along $z$. This is the desired situation for operating a LAr TPC. It is clarified by the two graphs shown in Figure~\ref{fig:GreinacherModel}. The dark blue curves correspond to a fully charged voltage multiplier and illustrate the case with $t\!\rightarrow\!\infty$ both for $U$ (top) and for $E_L$ (bottom). The parameter $t/\tau$ that appears in the exponential terms of Equations~\ref{eq:GreinModel} and~\ref{eq:GreinModelE} is the quantity describing the state to which the circuit is charged. Apart from the situation of a fully charged high voltage multiplier, given in dark blue, five intermediate charging states, reaching from $t/\tau = 0.1$ (red) to $t/\tau = 2.0$ (light blue), are shown in Figure~\ref{fig:GreinacherModel}. They clarify how far the longitudinal electric field is from being uniform in case the circuit is not fully charged.

\subsection{Drift field calibration}
From straight ionization tracks produced by UV laser beams one can extract information about the strength and direction of the electric field in the drift volume of the TPC. These tracks can be used to decouple the longitudinal from the transverse electric field components and to estimate the strength of the parasitic transverse field in ARGONTUBE. 

Given the longitudinal electric field, the drift time $t_d(z)$ needed for a test charge released at the drift coordinate $z$ to reach the readout plane ($z=0$), is obtained by integrating along the drift direction
\begin{equation}
  t_d(z) = \int_0^{z} \frac{1}{v_L(E_L(z'))} \,\,dz', \label{eq:intTd}
\end{equation}
where $v_L\left(E_L\left(z'\right)\right)$ denotes the longitudinal drift velocity in the interval $[z', z'+dz']$.

Equation~\ref{eq:intTd} combined with experimentally determined dependence of the drift velocity $v_L$ on the electric field strength~\cite{wal99} defines the drift time of the electrons produced at any point in the drift cage for a given longitudinal field distribution. The total drift time can be measured for electrons released from the cathode by photoelectric effect (see the charge cloud at the track end in Figure~\ref{fig:lastrack}). For derivation of the longitudinal field this is the only experimental constraint for the model. The $v_L(E)$ dependence from~\cite{wal99} is a monotonically increasing function of the electric field. The dependence of the electric field on time  $E_L(z, \,t)$ at any given $z$ is also a monotonically increasing function. Therefore, the integral \ref{eq:intTd} is, as well, a monotonic function of $t/\tau$. Hence, the constraint $t_{d}(z_c) = t_c$, where $t_c$ is the measured total drift time across the cathode-anode distance $z_c$, gives unambiguous solution for the only free parameter $t/\tau$.
This solution can be found numerically by successive binary division approximation for a given charging state of the Greinacher circuit. Since the measurement error of $t_c$ is negligibly small, the main contribution to the error on $t/\tau$  is given by uncertainty of $v_L(E)$  parametrization from~\cite{wal99}. The resulting error on the drift field $E_L(z)$ can be derived from \ref{eq:GreinModelE} by Gaussian error propagation.

With a typical cathode nominal voltage $U_\infty = 130\,$kV, a total drift length of $z_c = 4.76\,$m, and the measured total drift time $t_{d}(z_c) = 6.616\,$ms (compare Figure~\ref{fig:lastrack}), this value is found to be $t/\tau = 1.09 \pm 0.01$. Considering the distributions in Figure~\ref{fig:GreinacherModel}, it is clear that the longitudinal electric field is indeed highly non-uniform in ARGONTUBE.

The derived value for $t/\tau$ could, in principle, be compared to the expected one, derived from the parameters of the Greinacher circuit capacitors and resistors. Practically it is not done for two reasons. The first is the difficulty of taking into account all parasitic capacitances in the circuit. The second is related to the typical ARGONTUBE operation sequence. In order to reach required argon purity 
a time period of the order of a day is required. Estimation of argon purity is based on observation of tracks in the detector, so having non-zero drift field is crucial. During this time Greinacher circuit is charged in consecutive steps to avoid electrical discharges. There is no tool in ARGONTUBE to fully discharge the partly charged Greinacher circuit. Therefore, the charging time $t$ for estimation of $t/\tau$ can not be accurately measured. 

\begin{figure}[ht]
  \centering
  \includegraphics[width=1\textwidth]{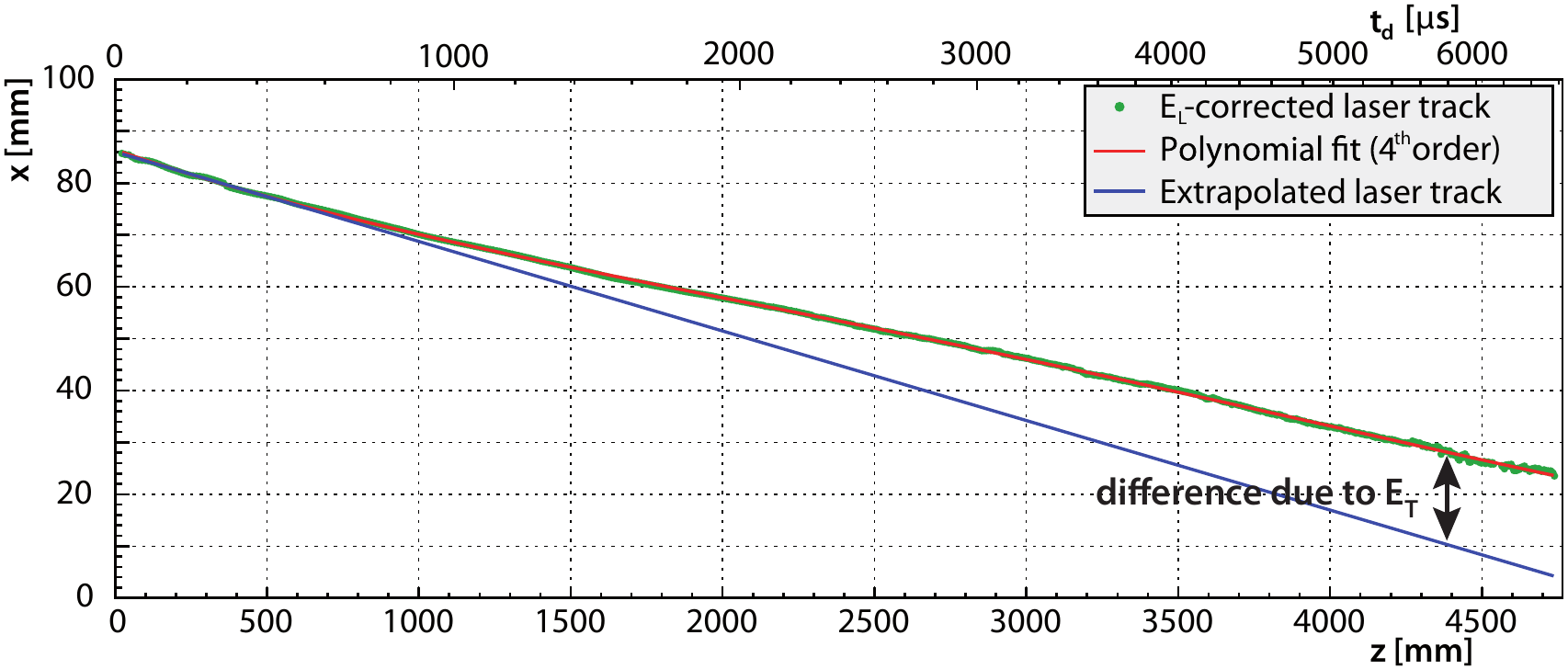}
  \caption[test]{The red line represents a polynomial fit to the laser track (green markers) shown in Figure~\ref{fig:lastrack} corrected for longitudinal disuniformities of the drift field using the Greinacher model and measurements described in the text. The blue line shows the extrapolated laser track determined by fitting the green markers with a straight line for $z \in [0, 300]\,$mm assuming $E_T(z) = 0$ in the specified range of $z$. The plot features two different scales along the abscissa, drift time $t_d$ (top axis) and drift coordinate $z$ (bottom axis) which are related by the function $t_d(z)$.}
  \label{fig:Elcorrected}
\end{figure}

Once the longitudinal coordinate of the laser track shown in Figure~\ref{fig:lastrack} is corrected using the inferred $E_L(z)$ distribution (red line in Figure~\ref{fig:Elcorrected}), the residual track curvature can be entirely attributed to the transverse field distortion $E_T(z)$. To determine the latter, one performs a straight line fit (shown in blue in Figure~\ref{fig:Elcorrected}) for $z \in [0, 300]\,$mm to the laser track corrected for $E_L(z)$ assuming that $E_T(z) = 0$ in the specified range of $z$. By analyzing the residual track deviation from the extrapolated laser track, one can determine the transverse drift velocity $v_{T}$ at each point along the laser-induced track and, therefore, derive the distribution of $E_T(z)$ in this specific region of the detector sensitive volume.


\begin{figure}[ht]
  \centering
  \includegraphics[width=1\textwidth]{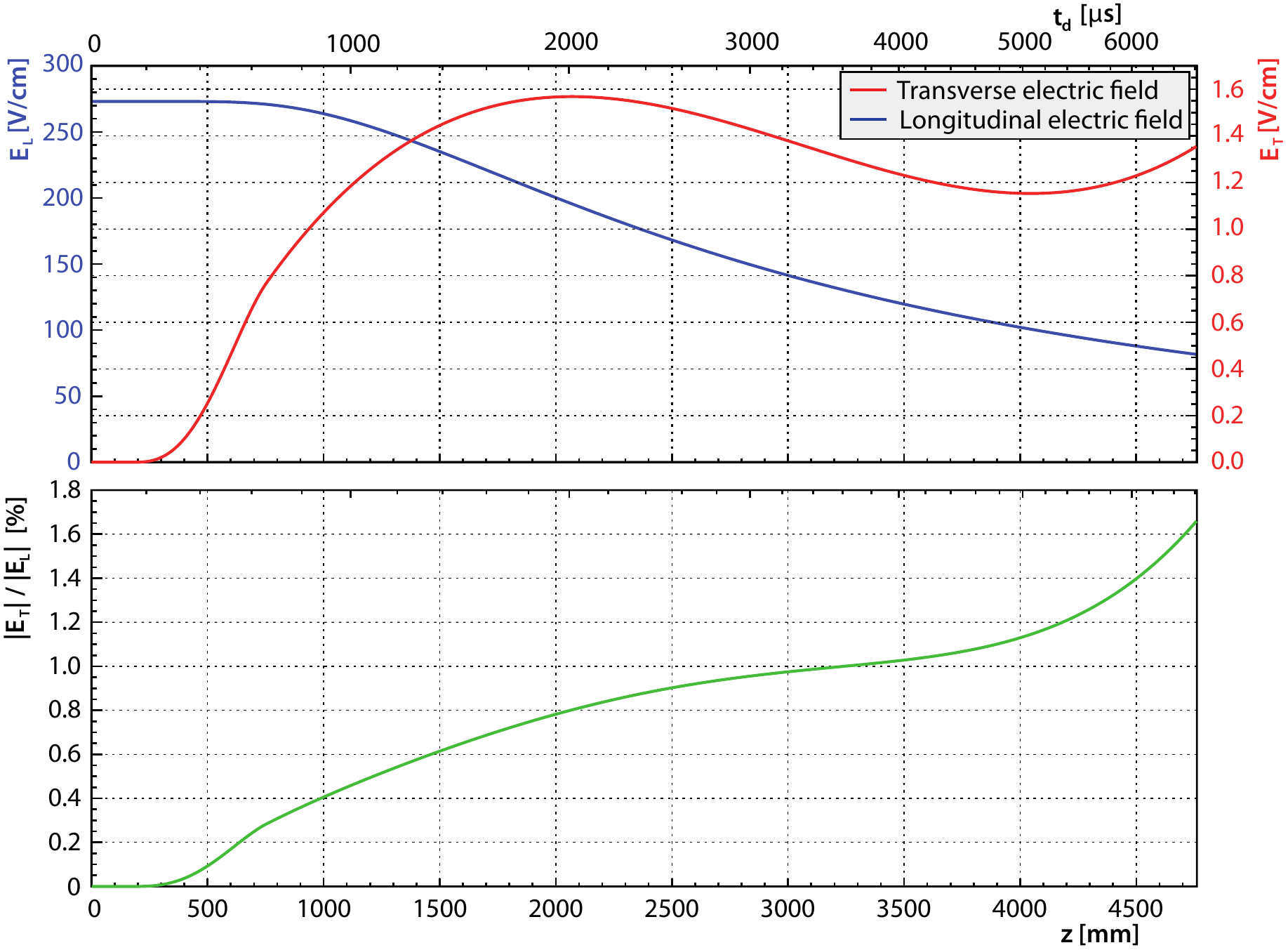}
  \caption{Top: the two electric field components in the ARGONTUBE detector drift volume. Bottom: Ratio of the absolute transverse to the longitudinal field strengths along the drift direction.}
  \label{fig:ElEt}
\end{figure}

The final result of the analysis is shown in Figure \ref{fig:ElEt}. The top plot represents the distribution of the longitudinal component (blue) and a projection of transverse component (red) of the drift field along the laser-induced track region. 
Deriving second projection in a similar way will result in full reconstruction of the $\vec{E}$ vector along the track. This possibility can not be realized on ARGONTUBE data because of the detector induction-collection charge readout scheme ~\cite{ARGONTUBE0,ARGONTUBE1,ARGONTUBE2}.
While the amplified signal from the charge collection plane used for this analysis is proportional to the input current in a wide bandwidth (1\%/ms of droop in response to a rectangular current pulse), the induction signal is proportional to its derivative on time. Therefore, in the situation with long track at a small angle w.r.t. drift direction, the induction plane signal does not allow to reconstruct second projection of laser-induced track in detail. This issue does not exist for detectors with a drift field transverse to a dominant track direction.
Full reconstruction of drift field vector would allow to cross-check the assumption of negligible contribution of the space charge by testing the condition $\nabla\cdot\vec{E}=0$.

While the longitudinal field component is valid for the whole sensitive volume of the detector, the transverse component is only correct along the laser path of the specific track and is found to vary within the detector volume, depending on the location w.r.t the Greinacher circuit components.
Steering the laser beam across the sensitive volume allows to reconstruct $E_T$ for several beam locations and hence to build up a full 3D map of the drift field components distribution.

The drift field data obtained for several runs of ARGONTUBE were used as a basis for the calibration of the detector response to the ionization charge (using the field-dependent charge recombination factor) and for charge diffusion measurements. These analyses will be presented in forthcoming publications.

\section{Conclusions}
The analysis of the apparent curvature of the ionization tracks in the ARGONTUBE LAr TPC produced with a narrow UV laser beam with $\lambda$=266~nm in combination with an analytic model of the Greinacher voltage multiplier circuit allows to decouple the longitudinal and the transverse components of the drift field. The distribution of the two components along the track volume can be effectively derived. This distribution sets the basis for further analyses of the detector response to ionization and of excess electron diffusion along the $\approx$5~m long drift in liquid argon.

\section{Acknowledgement}
We express our gratitude to engineering and technical staff in Bern for the design, manufacturing and assembly of the detector. 



\begin{thebibliography}{10}

\expandafter\ifx\csname url\endcsname\relax
\def\url#1{\texttt{#1}}\fi
\expandafter\ifx\csname urlprefix\endcsname\relax\def\urlprefix{URL }\fi









\bibitem{ARGONTUBE0}
  I.~Badhrees et al.,
  JINST 7 (2012) C02011.

\bibitem{ARGONTUBE1}
  A.~Ereditato et al.,
  JINST 8 (2013) P07002.

\bibitem{ARGONTUBE2}
  M.~Zeller et al.,
  Nucl.\ Instrum.\ Meth.\ A 718 (2013) 454.

\bibitem{Badhrees:2010zz}
  I.~Badhrees et al.,
  New J. Phys. 12 (2010) 113024.

\bibitem{Greinacher}
H. Greinacher, Physikal. Zeitsch., vol 15 (1914) 410.



\bibitem{wal99}
   W. Walkowiak, Nucl. Inst. and Meth. A 449 (2000) 288-294.

\end{thebibliography}
\end{document}